\documentclass{PoS}

\def\as{\alpha_s}
\def\lb#1{\if 1#1 \ln\beta \else \ln^#1\beta \fi}

\title{tops at NLO and NNLO}

\ShortTitle{tops at NLO and NNLO}

\author{\speaker{Michal Czakon}%
        \thanks{This work was supported by the Heisenberg Programme of
          the Deutsche Forschungsgemeinschaft.}\\ 
        Institut f\"ur Theoretische Physik E, RWTH Aachen University
        D-52056 Aachen, Germany \\
        E-mail: \email{mczakon@physik.rwth-aachen.de}}

\abstract{We briefly review the recent progress in the study of higher
  order corrections to hadron scattering processes involving top quark
  pairs in the final state. In particular, we discuss the Monte-Carlo
  simulation of the $pp \rightarrow t \bar{t} b \bar{b}$ process at
  the next-to-leading order with the help of the {\tt Helac-NLO}
  system, and the status of the soft-gluon resummation program for $pp
  \rightarrow t \bar{t}$ near threshold at the next-to-next-to-leading
  order.}

\FullConference{RADCOR 2009 - 9th International Symposium on Radiative Corrections (Applications of Quantum Field Theory to Phenomenology) \\
        October 25-30 2009\\
        Ascona, Switzerland}

\begin{document}


\section{Introduction}

Production of top quark pairs in hadronic collision, i.e. at the
TeVatron and LHC, is and will certainly remain an attractive subject
of study. From the multitude of problems, two correspond to the
current forefront of science: top pair production with two additional
partons at the next-to-leading order of QCD, and top pair production
with up to two partons at the next-to-next-to-leading order. Both
problems have generated interesting advances in theory. Below, we
shall summarize the current status.


\section{NLO}

In this proceedings contribution, we present the results obtained with
the help of the {\tt Helac-NLO} system. This software consists of the
following: {\tt Helac-Phegas} \cite{Kanaki:2000ey}, {\tt Helac-1Loop}
\cite{vanHameren:2009dr}, based on the OPP method and {\tt CutTools}
\cite{Ossola:2006us}, and finally {\tt Helac-Dipoles}
\cite{Czakon:2009ss}.

We consider the process $pp \rightarrow  t\bar{t} b\bar{b} + X$ at NLO
\cite{Bredenstein:2008zb,Bevilacqua:2009zn} at the LHC,
i.e. for  $\sqrt{s} = 14$ TeV. For the top-quark mass, renormalized in the
on-shell  scheme, we take the numerical value $m_t = 172.6 $ GeV. All other
QCD partons including b quarks are treated as massless particles. All  
final-state b quarks and gluons with pseudorapidity $|\eta| <5$ are 
recombined into jets with separation $\sqrt{\Delta\phi^2 +\Delta y^2} > D 
= 0.8$ in the rapidity-azimuthal-angle plane via the IR-safe
$k_T$-algorithm. Moreover, we impose the following additional cuts on
the transverse momenta and the rapidity of two recombined b-jets:
$p_{T,b} > 20$ GeV, $|y_b|< 2.5$. The outgoing (anti)top quarks are
neither affected by the jet algorithm nor by phase-space cuts.

We consistently use the CTEQ6  set of parton distribution functions 
(PDFs), i.e. we take CTEQ6L1 PDFs with a 1-loop running $\alpha_s$ in LO and CTEQ6M PDFs
with a 2-loop running $\alpha_s$ in NLO, but the suppressed contribution from
b quarks in the initial state has been neglected. The number of active
flavors is $N_F = 5$, and the respective QCD parameters are $\Lambda^{LO}_5 =
165$ MeV and  $\Lambda^{MS}_5 = 226$ MeV. In the renormalization of the strong
coupling constant, the top-quark loop in the gluon self-energy is subtracted at
zero momentum. In this scheme the running of $\alpha_s$ is generated solely by
the contributions of the light-quark and gluon loops. By default, we set the 
renormalization and factorization scales, $\mu_{R}$  and $\mu_F$, to the 
common value $\mu_0 =m_t$.

\begin{figure}[th]
\begin{center}
\includegraphics[width=0.4\textwidth]{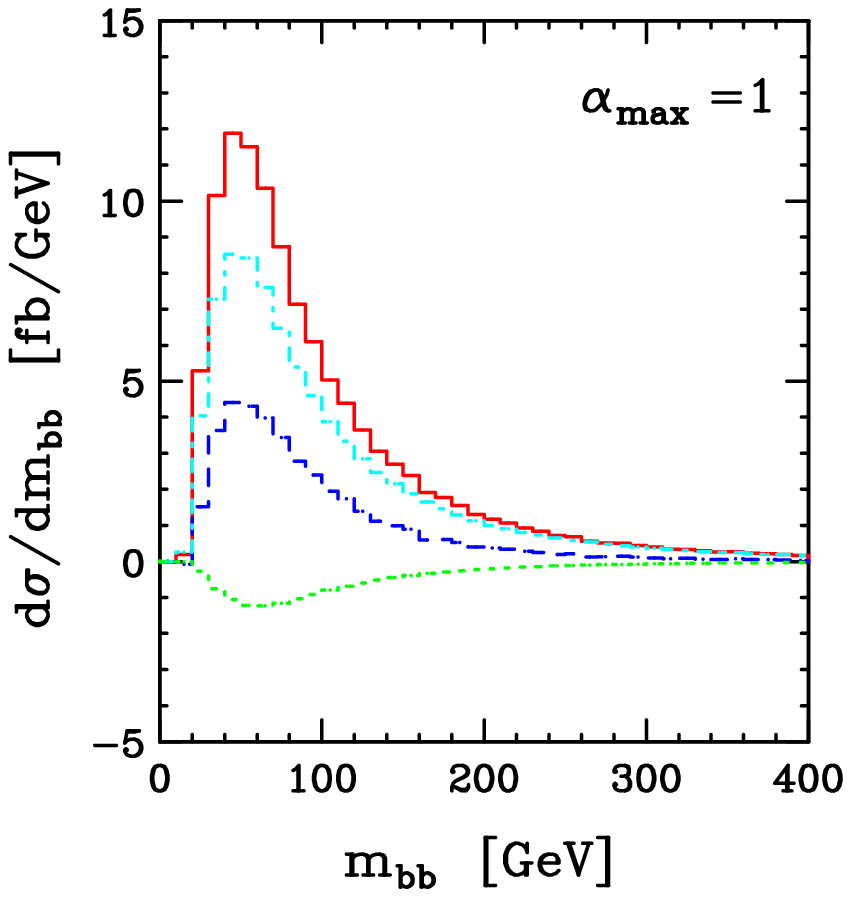}
\includegraphics[width=0.4\textwidth]{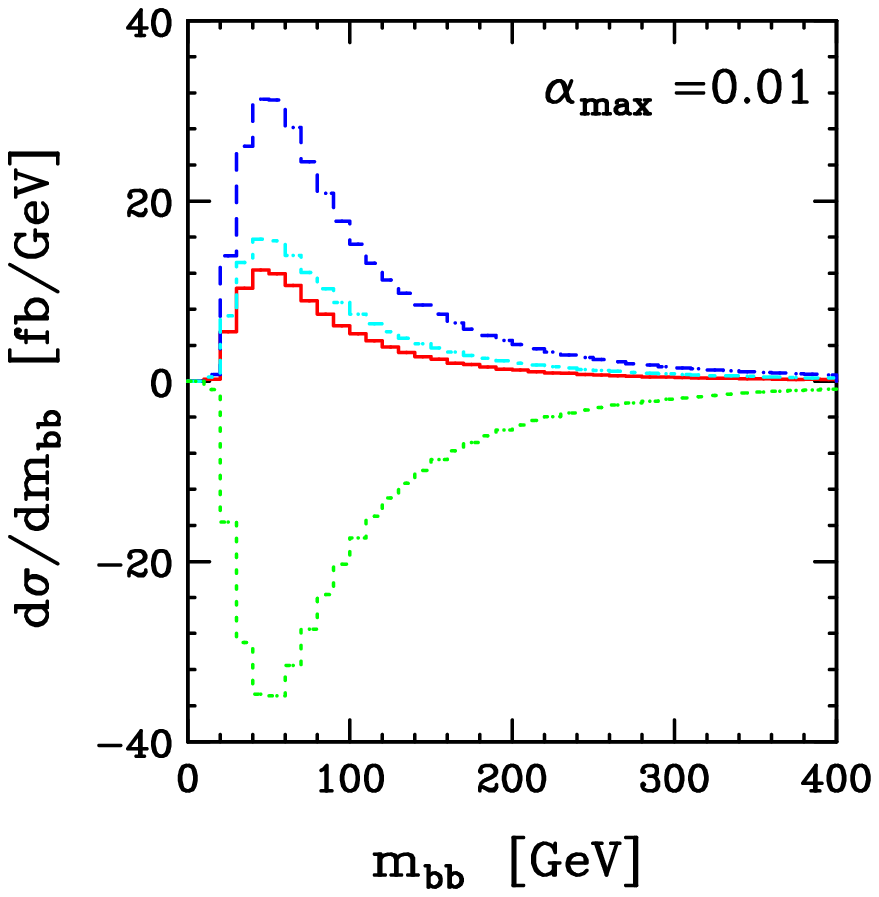}
\includegraphics[width=0.4\textwidth]{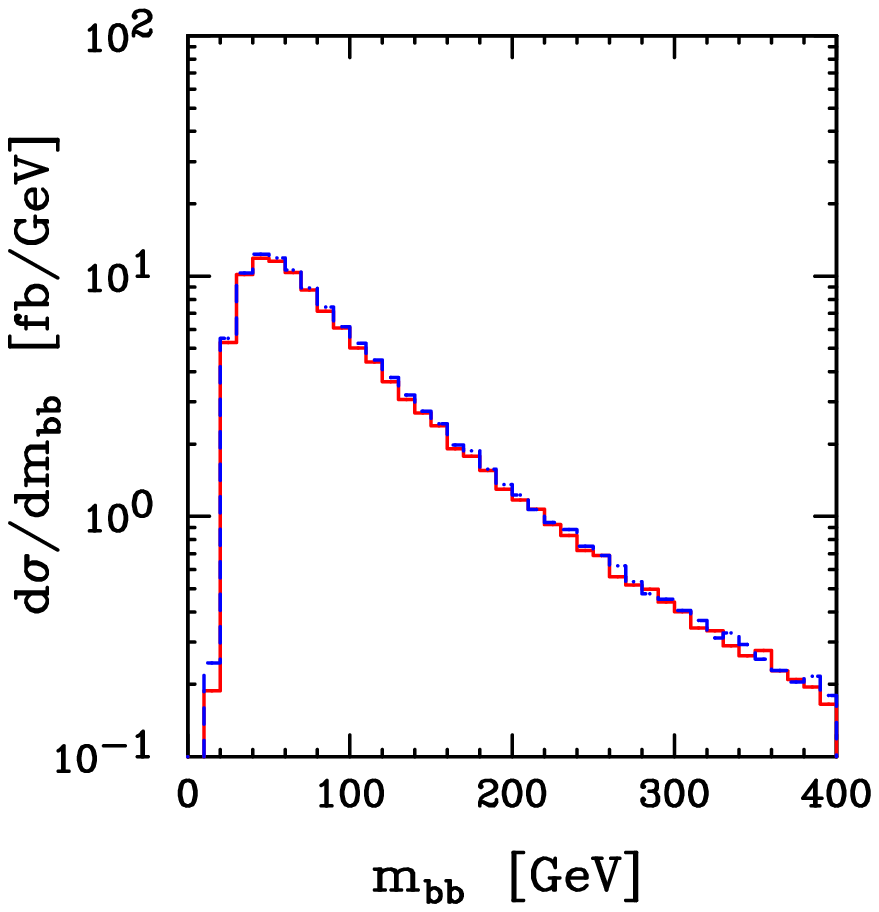}
\end{center}
\vspace{-0.2cm}
\caption{\it \label{fig:real} Distribution of the invariant mass
  $m_{b\bar{b}}$ of the bottom-anti-bottom pair for $pp\rightarrow
  t\bar{t}b\bar{b} + X$ at the LHC for different parts of the real
  radiation contribution with different choices of
  $\alpha_{max}$, $\alpha_{max} = 1$ and $\alpha_{max} = 0.01$. The red
  solid line corresponds to the sum of all contributions, the blue dashed
  line  represents the dipole subtracted real emission, the cyan
  dot-dashed line corresponds to the sum of the $K$ and $P$ insertion
  operators, and finally the green dotted line represents the $I$ insertion
  operator. The sum of all the contributions for the two different
  choices of $\alpha_{max}$ is depicted below.}
\end{figure}

\begin{figure}[t]
\begin{center}
\includegraphics[width=0.4\textwidth]{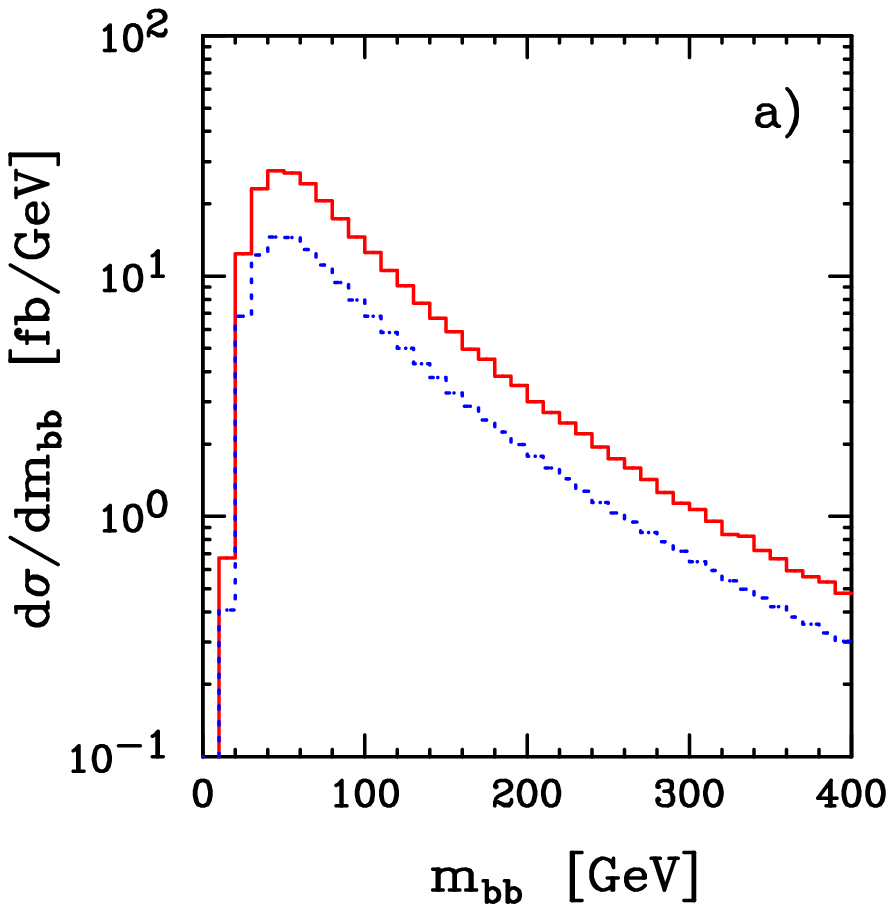}
\includegraphics[width=0.4\textwidth]{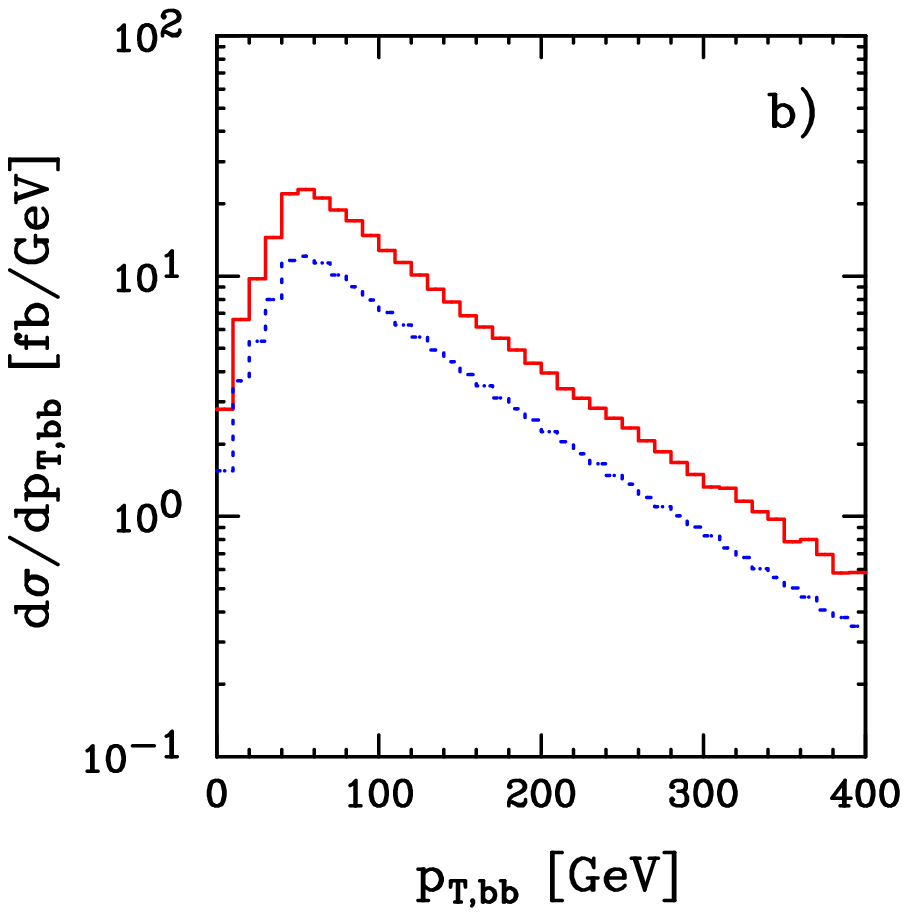}
\includegraphics[width=0.4\textwidth]{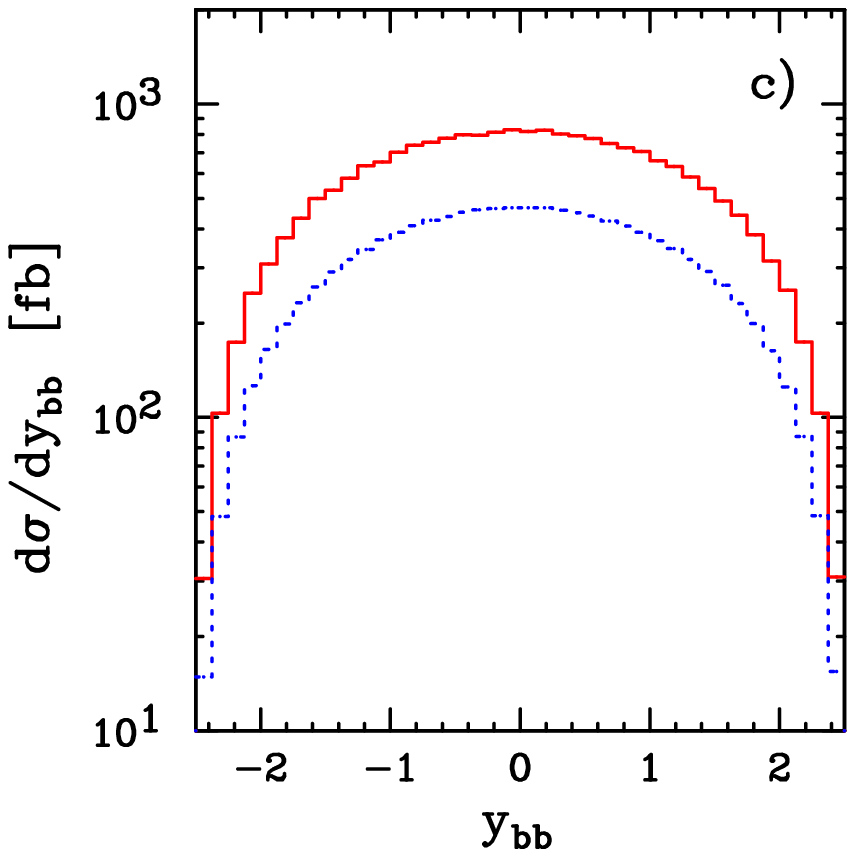}
\includegraphics[width=0.4\textwidth]{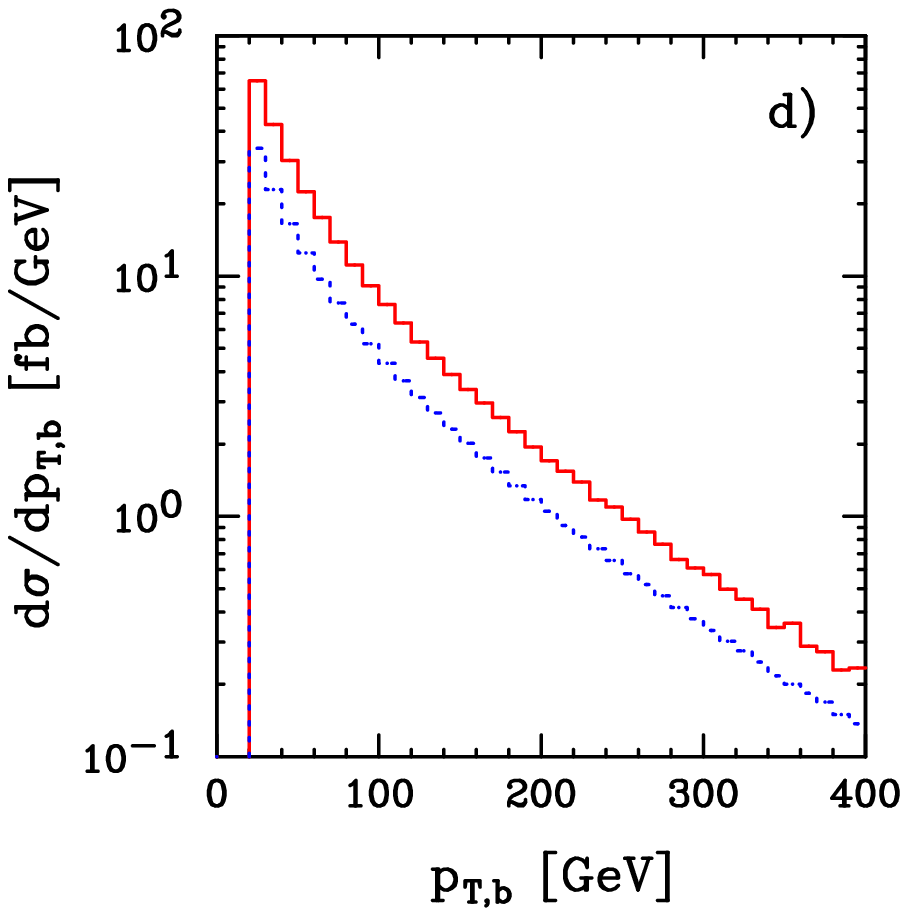}
\end{center}
\vspace{-0.2cm}
\caption{\it \label{fig:distros} Distribution of the invariant mass 
  $m_{b\bar{b}}$ of the bottom-anti-bottom pair (a), distribution in
  the transverse momentum  $p_{T_{b\bar{b}}}$  of the bottom-anti-bottom 
  pair (b), distribution in the rapidity $y_{b\bar{b}}$ of the
  bottom-anti-bottom pair (c)  and  distribution in the transverse momentum
  $p_{T_{b}}$  of the  bottom quark (d) for $pp\rightarrow t\bar{t}b\bar{b} 
  + X$ at the LHC at LO (blue dashed line) and NLO (red solid line). 
  All distributions have been obtained with $\alpha_{max}=0.01$. }
\end{figure}

With these parameters and assumptions we \cite{Bevilacqua:2009zn}
obtain the following results for the cross sections at LO and NLO,
which are in excellent agreement with \cite{Bredenstein:2008zb}
\begin{equation}
\sigma^{\mbox{{\scriptsize LO}}}_{t\bar{t}b\bar{b}}
(\mbox{LHC}, m_{t} = \mbox{176.2 GeV}, \mbox{CTEQ6L1})= 
{1489.2 ~}^{{\mbox{\scriptsize +1036.8 (70\%)}}}
_{-{\mbox{\scriptsize ~~565.8 (38\%)}}} ~~\mbox{fb} \; ,
\end{equation}
\begin{equation}
\sigma^{\mbox{{\scriptsize NLO}}}_{t\bar{t}b\bar{b}}
(\mbox{LHC}, m_{t} = \mbox{176.2 GeV}, \mbox{CTEQ6M})= 
{2636 ~}^{{\mbox{\scriptsize +862 (33\%)}}}
_{-{\mbox{\scriptsize 703 (27\%)}}} ~~\mbox{fb} \; .
\end{equation}
Besides the agreement for the cross sections, we were also interested
in distributions. As far as internal consistency checks are concerned,
we used the independence of the results on the size of the
subtraction phase space in the dipole formalism. This led to the plots
presented in Fig.~\ref{fig:real}. The notation and the definition of
the phase space cut-off parameter $\alpha_{max}$ is given in the
original publication.

The most important result of the study are the distributions presented
in Fig.~\ref{fig:distros}. They show the size of the differential K
factors, as well as the shape changes in the distributions
themselves. It should be noted that based on this complete
calculation, the NLO QCD corrections to arbitrary distributions can be
investigated within realistic selection cuts. On the other hand, this
calculation served as an important test of the system, as well as a
test of the correctness of the preceding results from
\cite{Bredenstein:2008zb}.


\section{NNLO}

The current status of fixed order NNLO calculations for top quark pair
production in hadronic collisions can be summarized as follows. The
most advanced evaluations concern the virtual corrections. In
particular, the latter have been derived in the high energy, fixed
angle limit in \cite{Czakon:2007ej}. Subsequently, the complete result
for the quark annihilation channel has been given numerically in
\cite{Czakon:2008zk}, whereas leading color and fermionic
contributions in the same channel are known from
\cite{Bonciani:2008az}. Additionally, the complete divergence structure
has been given in \cite{Ferroglia:2009ep}. Thus, the missing
contributions are the finite part of the virtual corrections in the
gluon fusion channel, and the real radiation contribution.

While the full calculation is still under way, there has been
important progress in the evaluation of the behavior at threshold. A
first, albeit incorrect, attempt at a complete NNLO threshold
expansion has been presented in \cite{Moch:2008qy}. However, it is only in
\cite{Beneke:2009ye} that the known contributions have been put together
with some missing ingredients due to potential interactions between
the quarks in order to obtain the correct and complete expansion.

Assuming that we decompose the cross section as follows
\begin{eqnarray}
\sigma_{ij}(\beta,\mu,m) &=& \sigma^{(0)}_{ij} \Bigg\{ 1
+ \frac{\as(\mu^2)}{4\pi} \left[\sigma^{(1)}_{ij}\right]
+\left(\frac{\as(\mu^2)}{4\pi}\right)^2
\left[\sigma^{(2)}_{ij}\right] + {\cal O}(\as^3)
\Bigg\} \; ,
\end{eqnarray}
where $i,j$ denotes the possible initial states (we concentrate only
on gluon fusion, gg, and quark annihilation, qq), and with
\begin{equation}
\beta = \sqrt{1-4m_t^2/s} \; ,
\end{equation}
we have (we have put the scales $\mu=m$)
\begin{eqnarray}
\label{eq:mainnnlo}
\sigma^{(2)}_{q\bar{q}} &=&
\frac{3.60774}{\beta^2}
+\frac{1}{\beta}\Big(-140.368\lb2+32.106\lb1+3.95105\Big) \nonumber \\ &&
+910.222\lb4-1315.53\lb3+592.292\lb2+528.557\lb1+C^{(2)}_{qq} \; ,
\nonumber \\[0.5cm]
\sigma^{(2)}_{gg} &=&
\frac{68.5471}{\beta^2}
+\frac{1}{\beta}\Big(496.3\lb2+321.137\lb1-8.62261\Big) \nonumber \\ &&
+4608\lb4-1894.91\lb3-912.349\lb2+2456.74\lb1+C^{(2)}_{gg} \; ,
\end{eqnarray}
Before we discuss the origin of these formulae, let us note that the
cross section expansion in the limit, where the emitted gluons are
soft (up to two at NNLO), but the final state is not at threshold, has
been given in \cite{Ahrens:2009uz}.

\begin{figure}[t]
\begin{center}
\includegraphics[width=.5\textwidth]{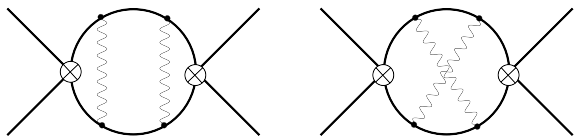}
\end{center}
\caption{\label{fig:color} Graphs relevant to the
  potential contributions in the singlet and octet
  channels discussed in the text. Crosses correspond to the singlet or
  octet color-projection operators.}
\end{figure}

The results in Eq.~(\ref{eq:mainnnlo}) require the following
ingredients: the two-loop soft anomalous dimension at threshold
\cite{Beneke:2009rj,Czakon:2009zw}, the matching coefficients at the
one-loop level in the two different color channels
\cite{Hagiwara:2008df} and finally the contribution of the potential
interactions between the heavy quarks. The latter can be derived from
threshold expansion of the top quark pair production cross section in
$e^+ e^-$ and $\gamma\gamma$ collisions \cite{Czarnecki:1997vz}. For
example the contribution for the gluon fusion is the same as that in
$\gamma\gamma$ scattering up to some minor modifications. If we take
the formula for the $R$ ratio in $\gamma$ collisions, we have
\begin{eqnarray}
R^{++}_S &=& 6 Q_t^4 N_c \beta 
\left ( 1 - \frac{\beta^2}{3} \right ) 
\cdot \left [ 1  + C_F \left ( \frac{\alpha_s}{\pi} \right ) 
\Delta^{(1)} 
+ C_F \left  ( \frac{\alpha_s}{\pi} \right )^2 \Delta^{(2)} \right],
\end{eqnarray}
where
\begin{eqnarray}
\Delta^{(2)} =&&
\hspace*{-5mm}
 C_F \Delta_{A} + C_A \Delta_{NA} + T_R N_L \Delta_L + 
T_R N_H \Delta_H,
\nonumber \\ \nonumber \\
\Delta_A &=& \frac{\pi^4}{12 \beta^2}
+ \left (- \frac{5}{2}+\frac{1}{8}\pi^2 \right ) \frac{\pi^2}{\beta}
+ \frac{27}{8}\pi^2+ \frac{25}{4}+ \frac{35}{192} \pi^4
- 2 \pi^2\ln( 2 \beta)+2 x_A;
\nonumber \\
 \Delta_{NA} &=& \left ( \frac{31}{72}
-\frac{11}{12} \ln(2 \beta) \right ) \frac{\pi^2}{\beta}
+\pi^2 \left ( \frac{5}{4} - \ln(2 \beta) \right ) 
+2 x_{NA};
\nonumber \\
 \Delta_{L} &=& \left ( -\frac{5}{18} + \frac{1}{3} 
\ln( 2 \beta) \right )\frac{\pi^2}{\beta} +2 x_L;
\nonumber \\
 \Delta_H &=& 2 x_H.
\end{eqnarray}
From this formula we need to first eliminate the effect of the
one-loop matching times the one-loop potential (Coulomb)
contribution. This is contained in $\Delta_A$ in the term proportional
to $1/\beta$. Next we retain only the terms with enhancements in
$\beta$, i.e. $1/\beta$ or $\log\beta$. Finally, we need to take into
account the possible color channels, since in $\gamma$ collisions
there is only a singlet contribution, whereas gluon fusion requires
also the octet. This is done by the replacement $C_F \rightarrow
C_F-C_A/2$. The correctness of the replacement can be checked by
considering the color structures depicted in Fig.~\ref{fig:color}

Ref.~\cite{Beneke:2009ye} contains a more extensive discussion of the
problem together with a proof that no other enhancements are present
at threshold at this order of the perturbative expansion.


\end{document}